\documentclass[aps,prl,twocolumn,groupedaddress]{revtex4}

\usepackage{graphicx}
\begin{document}
\title{Efficient Quasiparticle Evacuation in Superconducting Devices}
\author{Sukumar Rajauria$^{1,2,3}$}
\author{L. M. A. Pascal$^{1}$}
\author{Ph. Gandit$^{1}$}
\author{F. W. J. Hekking$^{4}$}
\author{B. Pannetier$^{1}$}
\author{H. Courtois$^{1}$}
\affiliation{$^{1}$Institut N\'eel, C.N.R.S. and Universit\'e Joseph
Fourier, 25 Avenue des Martyrs, B.P. 166, 38042 Grenoble, France}
\affiliation{$^{2}$Center for Nanoscale Science and Technology,
NIST, Gaithersburg, Maryland 20899, USA} \affiliation{$^{3}$Maryland
NanoCenter, University of Maryland, College Park, MD 20742, USA}
\affiliation{$^{4}$L.P.M.M.C., Universit\'e Joseph Fourier and CNRS,
25 Avenue des Martyrs, B.P. 166, 38042 Grenoble, France.}

\date{\today}
\begin{abstract}
We have studied the diffusion of excess quasiparticles in a
current-biased superconductor strip in proximity to a metallic trap
junction. In particular, we have measured accurately the
superconductor temperature at a near-gap injection voltage. By
analyzing our data quantitatively, we provide a full description of
the spatial distribution of excess quasiparticles in the
superconductor. We show that a metallic trap junction contributes
significantly to the evacuation of excess quasiparticles.
\end{abstract}


\maketitle

In a normal metal - insulator - superconductor (N-I-S) junction,
charge transport is mainly governed by quasiparticles \cite{Rowell}.
The presence of the superconducting energy gap $\Delta$ induces an
energy selectivity of quasiparticles tunneling out of the normal
metal \cite{nahum,giazotto}. The quasiparticle tunnel current is
thus accompanied by a heat transfer from the normal metal to the
superconductor that is maximum at a voltage bias just below the
superconducting gap (V $\leq \Delta$/e). For a double junction
geometry (S-I-N-I-S), electrons in the normal metal can typically
cool from 300 mK down to about 100 mK \cite{leivo, giazotto,
sukumar07PRL}. However, in all experiments so far the electronic
cooling is less efficient than expected \cite{leivo, sukumar07PRL}.
It has been proposed that this inefficiency is mostly linked to the
injected quasiparticles accumulating near the tunnel junction area.
This out-of-equilibrium electronic population, injected at an energy
above the superconductor energy gap $\Delta$, relaxes by slow
processes such as recombination and pair-breaking processes. The
accumulation of quasiparticles is aggravated in sub-micron devices,
where the relaxation processes are restricted by the physical
dimensions of the device, leading to an enhanced density of
quasiparticles close to the injection point. These quasiparticles
can thereafter tunnel back into the normal metal \cite{Jochum},
generating a parasitic power proportional to the bias current
\cite{sukumar08arxiv,VOutilainen}. The same phenomenon is relevant
to other superconductor-based devices such as qubits \cite{Lang},
single electron transistors \cite{Court} and low temperature
detectors \cite{Booth}.

In hybrid superconducting devices fabricated by multiple angle
evaporation, a normal metal strip in tunnel contact with the
superconducting electrode acts as a trap for excess near-gap
quasiparticles, which removes them from the superconductor. This
mechanism is usually not fully efficient due to the tunnel barrier
between the normal metal and the superconductor \cite{Pekolatrap}. A
detailed theory of non-equilibrium phenomena in a superconductor in
contact with normal metal traps has been developed
\cite{VOutilainen}. However, a quantitative comparison between
experiments and theoretical predictions is so far still missing.

\begin{figure}[h]
\centering
    \includegraphics[width=1\linewidth]{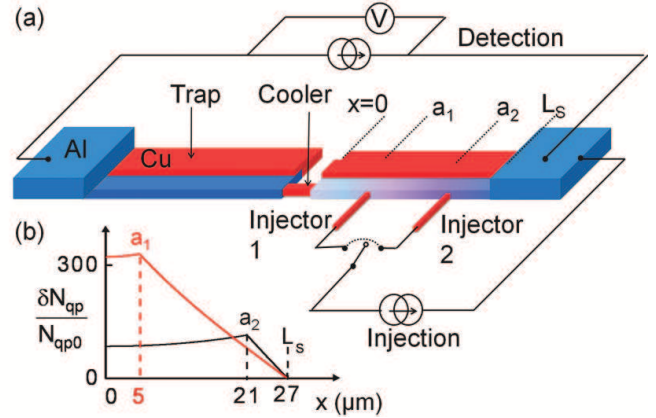}
    \caption{(a) Schematic of the sample design with a trap junction on each S-strip of the superconducting microcooler. The curve
    shows the spatial profile of the excess quasiparticles $\delta N_{qp}$ along the S-strip. The two injectors are located at a$_{1}$ = 5 $\mu$m or a$_{2}$ = 21 $\mu$m; the detector (cooler junction island) is at x = 0 $\mu$m and the length of the S-strip is $L_{S}$ = 27 $\mu$m. (b) Calculated spatial decay of excess quasiparticles density $\delta$N$_{qp}$ along the S-strip. The
    two injectors are biased at eV/$\Delta$=4.  }
    \label{fig:1}
\end{figure}

In this Letter, we present an experimental investigation of the
diffusion of out-of-equilibrium quasiparticles in a superconducting
strip covered with a trap junction. A N-metal is used to inject
quasiparticles in the S-strip. The local superconductor temperature
is inferred from the heating of the central N-island of a S-I-N-I-S
junction. We quantitatively compare our experimental data with a
recently discussed theoretical model \cite{sukumar08arxiv}.

We have used a S-I-N-I-S cooler device with a trap geometry similar
to the one studied in Ref. \cite{sukumar08PRL}, see Fig. 1. These
devices are fabricated using electron beam lithography, two-angle
shadow evaporation and lift-off on a silicon substrate having 500 nm
thick SiO$_{2}$ on it. The central normal metal Cu electrode is 0.3
$\mu$m wide, 0.05 $\mu$m thick and 4 $\mu$m long. The 27 $\mu$m long
symmetric S-strips of Al are then partially covered, through a
tunnel barrier, by a Cu strip acting as a trap junction. At their
extremity, the S-trips are connected to a contact pad acting as a
reservoir. In addition to the cooler island, we added two normal
metal Cu tunnel injector junctions of area around 0.09 $\mu$m$^2$ on
one S-strip. Injector 1 and 2 are at a distance of $a$ = 5 $\mu$m
and 21 $\mu$m respectively from the central Cu island. The Al tunnel
barrier is assumed to be identical in the cooler, probe and trap
junctions, since they have similar specific tunnel resistance. The
normal state resistance of the double-junction S-I-N-I-S cooler is
1.9 k$\Omega$. The normal state resistance of N-I-S injector
junctions 1 and 2 are respectively 2.5 k$\Omega$ and 2.3 k$\Omega$.
The diffusion coefficient of the Al S-strip film was measured at 4.2
K to be 30 cm$^{2}$/s.

N-I-S tunnel junctions are known to enable controlled quasiparticle
injection in a superconductor \cite{Tinkham,Yagi,Hubler}. The tunnel
current through a N-I-S junction is given by:
\begin{equation}
I(V)=
\frac{1}{eR_{N}}\int_{0}^{\infty}n_{S}(E)[f_{N}(E-eV)-f_{N}(E+eV)]dE
\label{eq:1}
\end{equation}
where $R_{N}$ is the normal state resistance, $f_{N}$ is the
electron energy distribution in the normal metal and $n_{S}$ is the
normalized density of states in the superconductor. In a
superconducting wire undergoing quasiparticle injection, the
superconductor gap $\Delta(T_{S})$ is suppressed locally. As this
gap can be extracted from a N-I-S junction current-voltage
characteristic, such a junction can be used for quasiparticles
detection. Usually, an effective superconductor temperature
$T_{\textrm{S}}$ is inferred from the superconductor gap-temperature
dependence. Fig. 2(a) displays the differential conductance $dI/dV$
of a N-I-S probe junction (similar to an injector in Fig. 1(a))
located on a S-strip at different cooler bias voltages. At high
injection, the gap suppression appears clearly, and enables a good
determination of the superconductor effective temperature. This
approach was used in numerous previous studies
\cite{Tinkham,Hubler,Yagi}.  At lower injection with a voltage
closer to the gap voltage, the tunnel characteristic becomes little
sensitive to quasiparticle injection. For instance, in Fig. 2a the
probe junction characteristic at 1 mV injection voltage almost
overlaps the equilibrium characteristic (at 0 mV). This limitation
comes naturally from the saturation of the superconducting gap at
low temperature $T_{\textrm{S}}$$\ll T_{\textrm{c}}$, where
$T_{\textrm{c}}$ is the superconductor critical temperature. So far,
the insensitivity of the N-I-S junction characteristic at low
injection bias has been a major roadblock in investigating the decay
of quasiparticles injected at energies just above the gap
\cite{Yagi,Hubler}.


Instead of measuring directly the superconductor, a better detection
sensitivity can be achieved by measuring the temperature of a small
N-island connected to superconductor through a tunnel barrier
\cite{Ullom}. In the absence of excess quasiparticles in the
superconductor, the N-island is in thermal equilibrium with it. When
the superconductor is under injection, some of the excess
quasiparticles population will escape by tunneling (even at zero
bias) from the superconductor to the central N-island. The injected
quasiparticles population will then reach a quasi-equilibrium in the
N-island with a electronic temperature $T_{\textrm{N}}$ different
from the cryostat temperature $T_{\textrm{bath}}$.


\begin{figure}[h]
\centering
    \includegraphics[height=0.68\linewidth, width=1\linewidth]{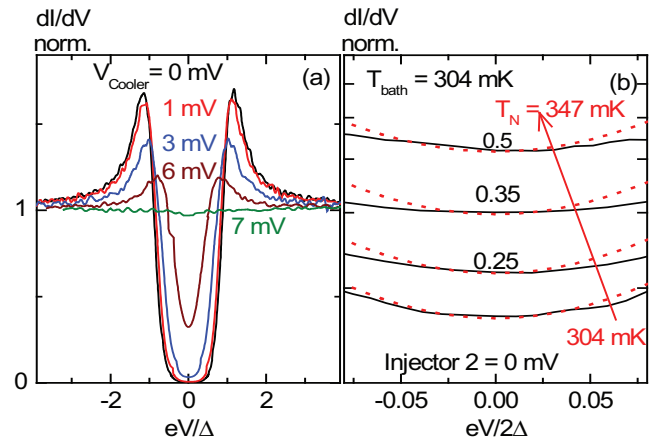}
    \caption{(a) Probe junction differential conductance under different injection bias voltage from the S-I-N-I-S cooler junction.
    (b) Low bias cooler junction data (solid black lines) at different injector 2 voltage level compared to calculated isotherms (red dashed lines) as obtained from Eq. 1 with $\Delta$ = 0.22 meV at different T$_{N}$.}
    \label{fig:1}
\end{figure}
In this study, N-I-S junctions located on one S-strip of a S-I-N-I-S
junction are used as quasiparticle injectors by current-biasing
them. This leads to a spatial distribution of the excess
quasiparticles density along the S-strip (see Fig 1 (b)). The
central N-metal in the S-I-N-I-S cooler geometry is used as a
detector for the quasiparticle density (at $x$ = 0) in the
superconductor. In the N-island, the phase coherence time of about
200 ps (measured from a weak localization experiment in a wire from
the same material) is much shorter than the mean escape time from
the island estimated to about 100 ns. The N-island electronic
population is then at quasi-equilibrium. Its temperature
$T_{\textrm{N}}$ can be extracted from the zero bias conductance
level of the S-I-N-I-S junction \cite{sukumar07PRL}. Further, the
superconductor temperature is inferred from the N-metal temperature
by considering its heat balance. As demonstrated below, this scheme
is highly sensitive down to about 200 mK, where the S-I-N-I-S
junction I-V becomes dominated by the Andreev current
\cite{sukumar08PRL}.

Fig. 2(b) shows the differential conductance of the cooler junction
(full black lines) at different injector-2 bias voltages along with
isotherms (dotted red lines) calculated from Eq. 1. Fig. 3(a)
displays the central N-metal temperature extracted from the
zero-bias conductance as a function of injector bias voltage. As the
injector bias increases above the bath temperature
$T_{\textrm{bath}}$, the temperature $T_{\textrm{N}}$ increases,
indicating that more quasiparticles tunnel from the S-trip to the
N-island.

In order to obtain the superconductor temperature $T_{\textrm{S}}$
at the cooler edge ($x$ = $0$), we need to consider the heat balance
in the normal metal. The heat flow across a N-I-S junction with
different quasiparticle distribution on either side of the tunnel
barrier is given by:
\begin{equation}
P_{heat}(T_{N},T_{S})=
\frac{1}{e^{2}R_{N}}\int_{-\infty}^{\infty}E n_{S}(E)[f_{N}(E)-f_{S}(E)]dE
\label{eq:2}
\end{equation}
where $f_{S}$ is the energy distribution function in the
superconductor at temperature $T_{\textrm{S}}$. It is compensated by
electron-phonon coupling power $P_{e-ph}$ so that $P_{heat}$  +
$P_{e-ph}$ = 0. Here, we have used the usual expression for the
electron-phonon coupling $P_{e-ph}=\Sigma U (T_{N}^{5}-T_{ph}^{5})$,
where $\Sigma$ = 2 $nW \cdot \mu m^{-3}\cdot K^{-5}$ in Cu is a
material-dependent constant and $U$ is the metal volume. For the
normal metal phonons, the electron-phonon coupling power is
compensated by the Kapitza power $P_{K}(T_{ph};T_{bath})=
KA(T_{bath}^{4} - T_{ph}^{4})$, where K is an interface-dependent
parameter and A the contact area. The inset of Fig. 3(a) displays
the correspondence between the superconductor temperature
$T_{\textrm{S}}$ and the central N-metal temperature
$T_{\textrm{N}}$. We took the fitted Kapitza coupling parameter
value $K.A$ = 144 $pW \cdot K^{-4}$ found in Ref.
\cite{sukumar08PRL} in a very similar sample. The grey area shows
the calculated temperature $T_{\textrm{S}}$ at different Kaptiza
coupling coefficient $K$ ranging from 120 $W \cdot m^{-2} \cdot
K^{-4}$ to infinity. The uncertainty in $T_{\textrm{S}}$ due to
uncertainty in $K$ is negligible below 400 mK and at
$T_{\textrm{S}}$ at 600 mK it is only 15 mK.

\begin{figure}[h]
\centering
    \includegraphics[height=0.65\linewidth, width=1\linewidth]{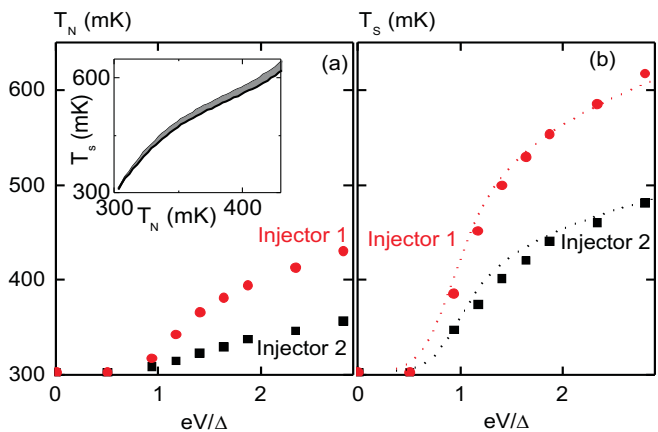}
    \caption{(a) Central N-metal electronic temperature T$_{N}$ dependence
on the injector 2 bias voltage at the bath temperature of 304 mK. Error in T$_{N}$ is smaller than the symbol size. Inset: calibration of the extracted superconductor temperature T$_{S}$
    to the measured normal metal temperature T$_{N}$. The grey area shows the uncertainty in T$_{S}$ for different values of the Kapitza coupling coefficient. (b) Corresponding extracted S-metal temperature T$_{S}$ at x = 0. Dotted lines are fits using D$_{qp}$ = 35 cm$^{2}$/s.}
    \label{fig:1}
\end{figure}

Fig. 3(b) and Fig. 4 show the extracted superconductor temperature
$T_{S}$ at the cooler edge ($x$ = 0) for different injection bias,
in the two injectors at three different bath temperatures
$T_{\textrm{bath}}$ = 100 mK, 304 mK and 500 mK respectively. We
have succeeded in obtaining accurately the superconductor
temperature for injection bias voltage close to the gap voltage. For
a given injection bias in the two injectors, the temperature
$T_{\textrm{S}}$ of the detector at $x$ = 0 is higher for a closer
injector. This confirms qualitatively the diffusion-based relaxation
of hot quasiparticles in the superconductor.

In an out-of-equilibrium superconductor, the quasiparticle density
$N_{qp}$ and the phonon density of $2\Delta$ energy $N_{2\Delta}$
are coupled to each other by the well-known Rothwarf-Taylor (R-T)
equations \cite{Rothwarf}. In a recent work \cite{sukumar08arxiv},
some of us have extended the R-T model to include the influence of
the trap junction on the quasiparticle diffusion. We considered a
superconducting strip covered by a second normal metal separated by
a tunnel barrier, which is in practice equivalent to a device
fabricated with a shadow evaporation technique \cite{giazotto}. The
normalized excess spatial quasiparticle density z(x) in the S-strip
is given by the solution of the differential equation:
\begin{eqnarray}
D_{qp}\frac{d^2z}{dx^{2}}=\frac{z}{\tau_{0}}+\frac{z+z^{2}/2}{\tau_{eff}}.
\end{eqnarray}
From this equation, one can find the quasiparticle decay length
$\lambda$ = $\sqrt{D_{qp}\tau_{eff}/\alpha}$, where $\tau_{eff}$ is
the material dependent effective recombination time and $\alpha$ = 1
+ $\tau_{eff}/\tau_{0}$ ($>$1) is the enhancement ratio of the
quasiparticle decay rate due to the presence of the trap junction.
When the trapping effect is dominant, the quasiparticle decay length
reduces to $\sqrt{D_{qp}\tau_{0}}$. The trap characteristic time
$\tau_{0}$ describes the rate of quasiparticles escaping to the
N-metal trap. It is defined as $\tau_{0} =
e^{2}R_{NN}N(E_{F})d_{S}$, where $R_{NN}$ is the specific resistance
of the trap junction and $d_{S}$ is the thickness of S. At
equilibrium, the density of quasiparticles $N_{qp0}$ in the
superconductor at temperature $T_{\textrm{S}}$ ($<T_{\textrm{c}}$)
decays exponentially: $N_{qp0}(T_{S})=N(E_{F})\Delta(\pi k
T_{S}/2\Delta)^{1/2}\exp[-\Delta/kT_{S}]$, where N$(E_{F})$ is the
density of states at the Fermi level.

In our experiment, the injectors are biased just above the gap
voltage. Here, we assume that the injected quasiparticles relax fast
(in comparison to other recombination processes) to the
superconductor gap energy level and thus can be afterwards
adequately described by the coupled R-T equations. To compare our
experimental result with the theoretical model, we have extended the
description shown in Ref. \cite{sukumar08arxiv}. We solved Eq. 3
numerically with boundary conditions so as to include the injection:
$\frac{dz}{dx}|_{+} - \frac{dz}{dx}|_{-} = \frac{I_{inj}}{\lambda}$
at x = a; the detection: $\frac{dz}{dx} = 0$ at $x$ = 0; and the
finite length of the S-strip: $z = 0$ at $x$ = L$_{S}$ into the
model. The last boundary condition $x$ = L$_{S}$ provides an
additional path for the excess quasiparticles to thermalize in
addition to the N-trap. Further, the locally enhanced quasiparticle
density $N_{qp}(x)$ is described by an equilibrium quasiparticle
density $N_{qp0}$ such that $N_{qp}(x=0) = N_{qp0}(T =T_{S})$ to
obtain the theoretical superconductor temperature $T_{\textrm{S}}$
\cite{Parker}.

\begin{figure}[h]
\centering
    \includegraphics[height=0.65\linewidth, width=1\linewidth]{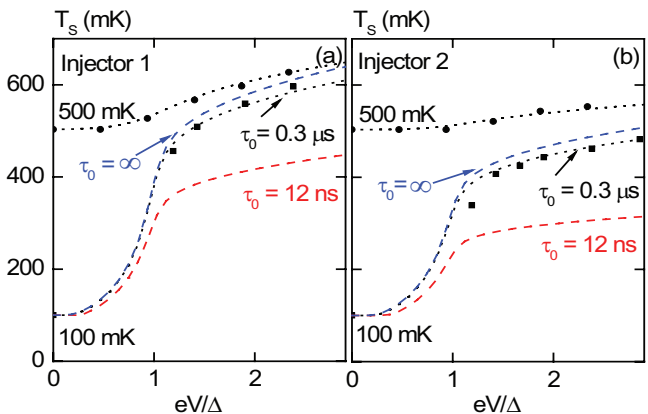}
    \caption{Extracted superconductor temperature T$_{S}$ as a function of injector bias voltage at a cryostat temperature of 100 mK (square dots) and 500 mK (circle dots). (a) and (b) correspond
to an injection from injector 1 and 2 respectively. Dotted lines show the fits with the parameter D$_{qp}$ = 35 cm$^{2}$/s. Red and blue dashed lines are the calculated curve with the same D$_{qp}$ = 35 cm$^{2}$/s  and for $\tau_{0}$ = 12 ns and $\infty$ respectively.}
    \label{fig:1}
\end{figure}
In the fit procedure, we used the calculated values of $\tau_{eff}$
and $\tau_{0}$. The calculated effective recombination time for Al
is $\tau_{eff}$ = 14 $\mu$s, \cite{kaplan} which is close to the
experimental value in Ref. \cite{Gray}. For our sample parameters,
the calculated trap characteristic time $\tau_{0}$ is equal to 0.3
$\mu$s. The model has then only one free parameter, which is the
quasiparticle diffusion coefficient $D_{qp}$. We obtain a
quantitative agreement between the theoretical predictions for
$D_{qp}$ = 35 cm$^{2}$/s and the experiment (dotted lines in Fig.
3(b) and Fig. 4) on the two injectors, for every injection voltage,
and for a bath temperature of 100 mK to 500 mK. The fit-derived
value of D$_{qp}$ is comparable to the measured diffusion
coefficient for Al at 4.2 K, and corresponds to $\lambda$ = 30
$\mu$m.

This excellent agreement demonstrates that we quantitatively
understand the diffusion of quasiparticles in the superconductor in
proximity with the metallic trap junction. The quasiparticles relax
mostly through two channels: N-metal trap (decay length $\lambda$)
and absorption in the reservoir (decay length L$_{S}$). In our
device, the decay length of both channels is around ~30 $\mu$m, thus
they act with a similar efficiency. Fig. 4 red and blue dashed lines
shows the calculated superconductor temperature $T_{\textrm{S}}$ at
$T_{\textrm{bath}}$ = 100 mK for a more transparent trap junction
$\tau_{0}$ = 12 ns and $\infty$, in parallel to the experiment and
its fit. For $\tau_{0} = \infty$, the excess quasiparticles relax at
the reservoir $x = L_{S}$. For $\tau_{0}$ = 12 ns, the decay length
is $\lambda$ = 6.5 $\mu$m, so that trapping then dominates the
quasiparticles absorption. At such transparency the influence of
proximity effect  cannot be ignored \cite{VOutilainen}.

In conclusion, we have  studied experimentally the diffusion of
quasiparticles injected in a superconductor with an energy close to
the gap voltage $\Delta$ in the presence of normal metal trap traps.
Our study demonstrates that in such devices quasiparticle trapping
competes with relaxation in the reservoir. This new knowledge is of
great importance in improving the geometry of the future cooling
devices and other superconductor, based low-temperature devices.

The authors are grateful to Nanofab-CNRS. SR acknowledges the
support of CNST NIST during preparation of manuscript. This work is
funded by MICROKELVIN, the EU FRP7 low temperature infrastructure
grant 228464.


\begin{thebibliography}{1}

      \bibitem{Rowell} J. M. Rowell and T. S. Sui, {Phys. Rev. B $\textbf{14}$, 2456 (1976).}

      \bibitem{giazotto} F. Giazotto, T. T. Heikkil\"{a}, A. Luukanen, A. M. Savin, and J. P. Pekola,  {Rev. Mod. Phys. $\textbf{78}$, 217 (2006).}

   \bibitem{nahum} M. Nahum, T. M. Eiles, and J. M. Martinis,  {Appl. Phys. Lett. $\textbf{65}$, 3123 (1994).}

   \bibitem{leivo} M. M. Leivo, J. P. Pekola, and D. V. Averin, {Appl. Phys. Lett. $\textbf{68}$, 1996 (1996).}


      \bibitem{sukumar07PRL} S. Rajauria, P. S. Luo, T. Fournier, F. W. Hekking, H. Courtois, and B. Pannetier,
 {Phys. Rev. Lett. $\textbf{99}$, 047004 (2007).
  }

    \bibitem{Jochum} J. Jochum, C. Mears, S. Golwala, B. Sadoulet, J. P. Castle, M. F. Cunningham, O. B. Drury, M. Frank, S. E. Labov, F. P. Lipschultz, H. Netel and B. Neuhauser, {J. Appl. Phys. $\textbf{83}$, 3217 (1998).}

      \bibitem{sukumar08arxiv} S. Rajauria, H. Courtois, and B. Pannetier, {Phys. Rev. B $\textbf{80}$, 214521 (2009).}

    \bibitem{Lang} K. M. Lang, S. Nam, J. Aumentado, C. Urbina and J. M. Martinis,  {IEEE Trans. on Appl. Superconduct. $\textbf{13}$, 2
    (2003)}; R. Lutchyn, L. Glazman and A. Larkin, {Phys. Rev. B $\textbf{72}$, 014517 (2008).}


    \bibitem{Court} N. A. Court, A. J. Ferguson, Roman Lutchyn and R. G. Clark, {Phys. Rev. B $\textbf{77}$, 100501(R) (2008).}

    \bibitem{Booth} N. E. Booth, {Appl. Phys. Lett. $\textbf{50}$, 2993
    (1987)}; C. M. Wilson and D. E. Prober, {Phys. Rev. B $\textbf{69}$, 094524 (2004).}


    \bibitem{Pekolatrap} J. P. Pekola, D. V. Anghel, T. I. Suppula, J. K. Suoknuuti,
A. J. Manninen, and M. Manninen, {Appl. Phys. Lett. $\textbf{76}$,
2782 (2000).}

    \bibitem{VOutilainen} J. Voutilainen, T. T. Heikkil\"{a} and N. B. Kopnin,  {Phys. Rev. B $\textbf{72}$, 054505 (2005)}; A.S. Vasenko and F. W. Hekking, {J. Low Temp Phys $\textbf{154}$, 221-232 (2009).}


      \bibitem{sukumar08PRL} S. Rajauria, P. Gandit, T. Fournier, F. W. Hekking, B. Pannetier and H. Courtois,
 {Phys. Rev. Lett. $\textbf{100}$, 207002 (2008).
  }


    \bibitem{Tinkham} M. Tinkham,  {Phys. Rev. B $\textbf{6}$, 1747 (1972).}

    \bibitem{Yagi} R. Yagi,  {Phys. Rev. B $\textbf{73}$, 134507
    (2006)}; K. Yu. Arutyunov, H. P. Auraneva and A. S. Vasenko, {Phys. Rev. B $\textbf{83}$, 104509 (2011).}



    \bibitem{Rothwarf} A. Rothwarth and B. N. Taylor,  {Phys. Rev. Lett. $\textbf{19}$, 61 (1967).}

    \bibitem{Parker} W. H. Parker,  {Phys. Rev. B $\textbf{12}$, 3667 (1975).}

    \bibitem{kaplan} S. B. Kaplan, C. C. Chi, D. N. Langenberg, J. J. Chang, S. Jafarey and D. J. Scalapino,  {Phys. Rev. B $\textbf{14}$,
    4854 (1976).}

    \bibitem{Gray} K. E. Gray,  {J. Phys. F: Met. Phys. $\textbf{1}$, 290 (1971).}


\bibitem{Ullom} J. N. Ullom, P. A. Fisher, and M. Nahum,  {Phys. Rev. B $\textbf{61}$, 14839 (2000).}

\bibitem{Hubler} F. Hubler, J. Camirand Lemyre, D. Beckmann and H. v. Lohneysen,  {Phys. Rev. B $\textbf{81}$, 184524 (2010).}

  \end{thebibliography}
 \end{document}